\documentstyle[preprint,eqsecnum,aps,prbbib]{revtex}
\tightenlines

\def\cm{~cm$^{-1}$}

\def\square{\kern1pt\vbox{\hrule height 0.6pt\hbox{\vrule width 
0.6pt\hskip 2pt \vbox{\vskip 4pt}\hskip 2pt\vrule width 0.6pt}\hrule
height 0.6pt}\kern1pt}
\def\h{_{\mathrm{H}}} 
\def\f{_{\mathrm{F}}} 
\def\xx{_{\mathrm{xx}}} 
\def\xy{_{\mathrm{xy}}}

\begin{document}

\title{Mid-infrared Hall effect in thin-film metals:\\
Probing the Fermi surface anisotropy in Au and Cu}

\author{J. \v Cerne, D.C. Schmadel, M. Grayson, G.S. Jenkins, J.R.
Simpson, and H.D. Drew}

\address{Department of Physics, University of Maryland, College Park, MD
20741, USA.}

\maketitle

\begin{abstract} 

A sensitive mid-infrared (MIR, 900-1100\cm, 112-136~meV) photo-elastic
polarization modulation technique is used to measure simultaneously Faraday
rotation and circular dichroism in thin metal films. These two quantities 
determine the complex AC Hall conductivity. This novel technique is applied
to study Au and Cu thin films
at temperatures down to 20~K and magnetic fields up to 8~T. The Hall
frequency $\omega\h$ is consistent with band theory
predictions.  We report the first 
measurement of the MIR Hall scattering rate
$\gamma\h$, which is significantly 
lower than that derived from Drude analysis of
zero magnetic field MIR transmission measurements. This difference is
qualitatively explained in terms of the anisotropy of the Fermi surface in
Au and Cu.


\end{abstract}


\date{}

\narrowtext

\section{Introduction}

The DC Hall effect is a standard tool for the study of the electronic
properties of conducting materials. In the high field limit
($\omega_c\tau >>1$, where $\omega_c$ is the cyclotron
frequency and $\tau$ is the carrier scattering time) it can can be shown
that the
Hall coefficient $R\h$ ($R\h\propto1/n$) can give the
number density $n$ of carriers. In many novel electronic materials, however,
because of the combination of large effective 
masses $m$ ($\omega_c\propto\frac{1}{m}$)  
and short defect
induced carrier scattering times $\tau$, it is 
not possible to achieve the high field
limit of the Hall effect. Under these conditions the Hall effect is
sensitive to the defect scattering in the sample, which complicates its
interpretation. Also, in many of these interesting materials, electron
interactions are strong and the Fermi liquid theory conditions that are
assumed in transport theory are possibly not met.  Nevertheless, the DC
Hall effect has provided interesting and important information on these
materials.\cite{dchall} The AC Hall effect offers the possibility of
overcoming the limitations of the DC Hall effect while providing
additional information on the electronic structure of materials. While DC
transport can be shown to be mainly sensitive to the mean free path of the
carriers,\cite{ong} AC transport is sensitive to the energy scales of the
system: the plasma frequency, the cyclotron frequency and the carrier
relaxation rates. At sufficiently high frequencies $\omega$ ($\omega \tau
>>1)$ the AC conductivity becomes insensitive to the impurity scattering,
thereby giving information about the electronic structure of the ``clean''
system. It can provide insight into the physics of many systems
ranging from conventional Fermi liquid metals to more exotic metals such as
high temperature superconductors\cite{htsc}  and 
magnetic transition metal oxides.\cite{cmr} 

Within Fermi liquid theory and the relaxation time approximation, and
assuming cubic symmetry, the
conductivity tensor can be expressed as 
integrals over the Fermi surface (FS).\cite{general}

\begin{mathletters}
\begin{eqnarray}
\sigma\xx=&&  \frac{e^2}{(2\pi)^3\hbar}
\oint_{\hbox{\tiny FS}} dS\ |{\bf v}(k)|\tilde\tau(k)
\label{eq;fsa}\\ 
\sigma\xy=&& \frac {e^3H}{(2\pi)^3\hbar^2c} 
\oint_{\hbox{\tiny FS}} dS\ {\bf e_z}\cdot\biggl[{\bf v}(k)\tilde\tau(k)\times 
{ {d[{\bf v}(k)\tilde\tau(k)]}\over{dk}}\biggr] 
\label{eq;fsb}
\end{eqnarray}
\end{mathletters}

\noindent where ${\bf v}$ is the carrier velocity, $k$ is the carrier
momentum, $\tilde\tau=\tau/(1-i\omega\tau)$ where $\tau$ is the
scattering time at point $k$ on the FS, $H$ is
the magnetic field, $e$ is the electron charge, $\hbar$ is Planck's
constant, $c$ is the speed of light, and ${\bf e_z}$ is the unit vector
along the $z$-axis (${\bf B}\parallel{\bf z}$).  Since $\sigma\xy$ 
depends on the cross product 
${\bf v}\times {\bf dv}$,
flat regions of the FS (where ${\bf v}\parallel{\bf dv}$) do not contribute
to $\sigma\xy$ whereas high curvature regions, which can produce large
angular differences between ${\bf v}$ and ${\bf dv}$, will be heavily weighted
in the integral in Eq.~\ref{eq;fsb}.

In the study of the AC Hall effect the 
complex Hall angle $\theta\h$ is
a particularly useful quantity. $\theta\h$ is defined as

\begin{equation}
\tan\theta\h=\frac{\sigma\xy}{\sigma\xx},
\label{eq;hallang0}
\end{equation}

\noindent where $\sigma\xx$ and $\sigma\xy$ are the diagonal and
off-diagonal components of the complex 
magneto-conductivity tensor.  Note that $\tan\theta\h$ is independent of
film thickness $d$, which is useful since $d$ may not be accurately known.
For a simple (Drude) metal, $\tan\theta\h$ reduces to 
$\omega _{c}\tau$ in the DC
limit. $\theta\h$ has proven to be especially interesting in high temperature 
superconductors where the scattering rate associated with $\theta\h$
shows striking qualitative and quantitative differences from the rate
associated with $\sigma\xx$. This behavior has been cited as evidence
for non-Drude and even non-Fermi liquid physics for high temperature
superconductors in the normal state.\cite{zhel,ioffe,anderson} Interesting
differences in scattering rates can also be found in more conventional
materials such as Au and Cu, as will be explored in this paper.

In general $\tan \theta\h$, as the ratio of two response functions, is
a complicated function which does not have a simple closed form. The
simplest generalization of $\theta\h$ to finite frequency
is:\cite{achall}

\begin{equation}
\tan\theta\h=\frac{\omega\h}{\gamma\h- i\omega}\approx\theta\h,  
\label{eq;hallang1}
\end{equation}

\noindent where $\omega\h$ is the Hall frequency and $\gamma\h$ is
the Hall scattering frequency.  In this experiment, since $\omega\h$ is small 
compared to $\omega$ and $\gamma\h$, we
will use the approximation $\tan\theta\h\approx\theta\h$ throughout this
paper.  Equation~\ref{eq;hallang1} is valid for a
Drude metal in which case, $\omega\h=\omega_{\mathrm{c}}$ and 
$\gamma\h=\gamma\xx$, where $\omega_{\mathrm{c}}$ and $\gamma\xx$ are the
conventional cyclotron frequency and isotropic Drude scattering rate,
respectively. Equation~\ref{eq;hallang1} is 
also valid for Fermi liquids for the
case of a $k$ independent scattering time.  Furthermore, it is the form obtained
in several proposed models of the normal state transport in high temperature
superconductors.\cite{zhel,ioffe,anderson}

Though $\sigma\xy$ and $\theta\h$ tend to be small for metals in the 
MIR (900-1100\cm), there
are a number of advantages in performing these higher frequency
measurements. First, the high frequency  allows
one to avoid impurity scattering or grain boundary effects which may
dominate lower frequency measurements. This is especially important in new
materials which often contain many impurities. Thus,
the MIR measurements can probe the intrinsic optical properties more
directly. Furthermore, the MIR measurements allow one to check the trends
observed at lower frequencies. Since $\theta\h$ obeys a sum rule (see 
Eq.~\ref{eq;sum}),\cite{drew} it is
very useful to be able to integrate $\theta\h$ to higher frequencies to
verify whether (and where) the Hall angle sum rule saturates or whether
there is more relevant physics at even 
higher frequencies.  Finally, since the high
frequency behavior of $\theta\h$ is constrained by the general
requirements of response functions, the asymptotic form for $\theta\h$ in
Eqs.~\ref{eq;hallang1} and \ref{eq;thetah2} becomes more accurate at 
higher frequencies.

In this paper, we examine the the MIR Hall effect in Au
and Cu thin films. We introduce a sensitive photo-elastic polarization modulation
technique  in Section~\ref{sec;polar}.
Section~\ref{sec;samples} describes the samples. Section~\ref{sec;results}
presents MIR magneto-optic
transmission measurements from 900\cm\ to 1100\cm.  We determine the complex
conductivity tensor $\sigma$ at temperatures down to 20~K and
magnetic fields up to 8~T. These results suggest that the Hall frequency
$\omega\h$ is in good agreement with band theory while the scattering rate
$\gamma\h$ determined from the Hall angle is significantly lower than the 
scattering rate $\gamma\xx$ measured in zero magnetic field transmittance
measurements. 
Section~\ref{sec;aniso} presents a qualitative model that is 
used to discuss these results. The anisotropy
of the FS of Au and Cu can explain the differences between the
magneto-optic and zero-field scattering rates.  Section~\ref{sec;irdc}
compares the results for the MIR
measurements with DC results. 
Appendix~\ref{app} provides theoretical background for the expression of
$\theta\h$ that is used to analyze the experimental data.

\section{Experiment}

\label{sec;experiment}

\subsection{Polarimetry Measurements}

\label{sec;polar}

Since $\omega\h<<(\gamma\h,\omega$) and since at high
freqencies Re$[\theta\h]\propto\omega^{-2}$ and
Im$[\theta\h]\propto\omega^{-1}$ (see Eq.~\ref{eq;hallang1}),
the MIR $\theta\h$ in metals is small, on the order of $10^{-3}$ Rad.
Therefore, a sensitive technique is required for the MIR $\theta\h$
measurements.   The MIR data are a direct measurement of the 
Faraday angle $\theta\f$, which is
the optical analogue of the $\theta\h$.  $\theta\f$ is defined as

\begin{equation}
\theta\f=\frac{t\xy}{t\xx}
\end{equation}

\noindent where $t\xx$ and $t\xy$ are the complex transmission
amplitudes.  In the thin film approximation, the relationship
between $\theta\f$ and $\theta\h$ is given by:

\begin{equation}
\theta\f \approx\tan \theta\f=\frac{t\xy}{t\xx}\approx
\frac{\frac{4nZ\sigma\xy}{(1+Z\sigma\xx)^2}}
{\frac{4n}{1+Z\sigma\xx}  }
=\biggl(1+\frac{1}{Z\sigma\xx}\biggr)\theta\h\hskip .5in;\hskip .5in
Z=\frac{Z_0d}{n+1},  
\label{eq;hallfar}
\end{equation}

\noindent where $Z_0$ is the impedance of free space, $n$ is the
substrate index of refraction, and $d$ is the film thickness.  As with 
$\theta\h$, the small
angle approximation applies to $\theta\f$ in this experiment, so
$\tan\theta\f\approx \theta\f$.  The experimental data was analyzed using
finite thickness film calculations, which deviated from the thin film
results by less than 10~\%.  Furthermore, since the 
conductances ($\sigma\xx d$) of Au and
Cu films are large, ${\frac{1}{{Z\sigma\xx}}}<<1$ so 
$\theta\h\approx \theta\f$. However, in this paper, $\sigma\xx$ is
determined through zero magnetic field transmission measurements, and
these corrections (less than 15~\%) are included in calculating $\theta\h$
and $\sigma\xy$ from $\theta\f$.

Figure~\ref{fig;setup} shows the experimental setup for measuring $\theta\f$.
A CO$_2$ laser produces linearly polarized MIR radiation (9-11 $\mu$m,
1100-900\cm, 112-136~meV). First, a conventional optical chopper modulates this
radiation at $\omega_0$ (80-150 Hz). In this schematic,
the laser polarization $P_L$ is along $\hat x$. In the Faraday geometry, the
radiation then passes through the sample which is located at the 
center of an 8~T
magneto-optical cryostat. In order to sensitively measure both the real and
imaginary parts of $\theta\f$, the radiation that is transmitted by the sample
is analyzed using a photoelastic modulator (PEM).\cite{hinds} The PEM
periodically  retards the phase of one linear
polarization component $E_x \hat x$ with respect to the orthogonal
component $E_y \hat y$ as follows:

\begin{equation}
\hbox{PEM}(E_x \hat x+ E_y \hat y)= E_x e^{i\Delta(t)} \hat x
+ E_y \hat y.
\end{equation}

\noindent $\Delta(t)$ is the sinusoidal phase modulation of $E_x \hat x$
with respect to $E_y \hat y$, and is given by:

\begin{equation}
\Delta(t)=\beta\cos(\omega_M t),
\end{equation}

\noindent where $\beta$ is the phase modulation amplitude and $\omega_M$
is the PEM modulation frequency (50~kHz). Since the sample is
axially symmetric along $B$, the transmittance tensor is diagonal when
represented in the circular polarization basis.  Therefore, changes in the incident
polarization only depend on: (1) the  relative difference in the phase of left
versus right circularly polarized light due to Re[$\theta\f$], 
which leads to a rotation (Faraday
rotation, FR) in the linearly polarized incident light; and (2) the relative
difference in the transmission of left versus right circularly polarized 
light due to Im[$\theta\f$],
which introduces ellipticity (circular dichroism, CD) to the linearly polarized
incident light.  The optical axis of the PEM is oriented parallel to that of the
laser radiation along $\hat x$ , so that no modulation occurs unless the sample
produces a $\hat y$-component in the 
polarization by either rotating the polarization (FR)  or introducing ellipticity to
the polarization (CD).  Finally, a static linear polarizer
$P_a$ selects the component of the radiation at 45$^{\circ}$ to $\hat x$. 
A liquid nitrogen-cooled mercury-cadmium-telluride (MCT) element detects the radiation, 
and three lock-in amplifiers
demodulate the resulting time-dependent signal.  Combining a bright source such as a CO$_2$ laser with
a sensitive MCT detector provides a signal to detector noise level of up to
$10^5$.  The high sensitivity is especially important due to the low
transmittance ($<3$~\%) of the samples used in this experiment.  The signal
intensity was kept within the linear response regime of the MCT detector. 

The FR (Re[$\theta\f$]) and
CD (Im[$\theta\f$] ) signals are related to the even and odd harmonics
of $\omega_M$, respectively. These harmonic signals can be normalized by
the average signal chopped at $\omega_0$ to obtain, for small $\theta\f$:

\begin{mathletters}
\begin{eqnarray}
{\frac{{I_{2\omega_M}}}{{I_{\omega_0}}}}=&&
\frac{4J_2(\beta)\hbox{Re}[\theta\f]}{1+|\theta\f|^2-2\hbox{Re}[\theta\f]J_0(\beta)}
\label{eq;far2w} \\
{\frac{{I_{3\omega_M}}}{{I_{\omega_0}}}}=&&
\frac{4J_3(\beta)\hbox{Im}[\theta\f]}{1+|\theta\f|^2-2\hbox{Re}[\theta\f]J_0(\beta)},
\label{eq;far3w}
\end{eqnarray}
\end{mathletters}

\noindent where $I_{n\omega_M}$, $J_n(\beta)$, $I_{\omega_0}$ are the
intensity of the $n$th harmonic of $\omega_M$, the $n$th order Bessel
function, and the average intensity chopped at $\omega_0$, respectively.  
Since $|\theta\f|^2<<1$ and $\beta$ is chosen so that $J_0(\beta)=0$, the
denominators in Eqs.~\ref{eq;far2w} and \ref{eq;far3w} are unity. The
$3\omega_M$ harmonic signal was chosen in Eq.~\ref{eq;far3w} over the
fundamental frequency in order to avoid background signals (such as
electrical pickup and interference modulation) that occur at $\omega_M$.
With this technique one can simultaneously measure both the real and
imaginary parts of $\theta\f$ with a sensitivity of approximately 1 part
in $10^4$ and $4\times 10^3$, respectively. The difference in sensitivity
for the real and imaginary parts of $\theta\f$ is mainly due to the fact
that at a typical PEM retardance $\beta\approx 2.39$~radian,
$J_2(\beta)\approx2\times J_3(\beta)$. The great stability of the
measurement is due in part to using a single detector to measure all the
signals simultaneously, so that detector and source drift can be accurately
normalized out.

Though all the parameters in Eqs.~\ref{eq;far2w} and \ref{eq;far3w} are
measured independently, the calibration of the system is verified by
removing the sample and rotating a quarter waveplate in front of the 
optical magnet shown in Fig.~\ref{fig;setup}. The signals as a 
function of quarter waveplate orientation angle are consistent with 
predictions from 
the initial calibration. For
more details on this measurement technique, see Ref.~\onlinecite{instr}.

\subsection{Samples}

\label{sec;samples}

The samples consisted of thin Au and Cu films grown on semiconductor
substrates using conventional vacuum thermal deposition techniques. The
film thicknesses were on the order of 10~nm, corresponding to DC
resistances in the range of $R_{\hbox{\square}}\approx6-10\ \Omega$. The
thickness was chosen to maximize the quality of the film and the
magneto-optic signals while maintaining a transmittance of approximately
5~\% at 1000\cm. Substrates consisted of 0.5~mm thick insulating Si 
($\rho\ge2000\ \Omega-$cm) or GaAs.  The DC residual resistance ratio 
($R_{\hbox{\tiny 300 K}}/R_{\hbox{\tiny 10 K}}$) for
the Au and Cu films is 1.6 and 1.3, respectively.  The low temperature
resistance is dominated by interface scattering, as will be shown in
Section~\ref{sec;aniso}.  Since interference (etalon) effects associated with the
substrate can
have a strong effect on FR and CD measurements, the substrates were either wedged
1 degree to remove multiply reflected beams or coated with a NiCr
broadband antireflection coating.\cite{ar} An antireflection coating with
$R_{\hbox{\square}}=157~\Omega$ for Si and $R_{\hbox{\square}}=
149~\Omega$ for GaAs reduced the etalon interference fringes to less than
5~\% of the transmittance signal.

\section{Results}

\label{sec;results}

Figure~\ref{fig;signal} plots Re$[\theta\f]$ and Im$[\theta\f]$ (see
Eqs.~\ref{eq;far2w} and \ref{eq;far3w}) as a function of magnetic field
$B$ at room temperature for a Cu sample. The MIR radiation frequency is
949\cm.  The substrate's background contribution to Re$[\theta\f]$ has been
removed.  Though both Si and GaAs substrates produced signicant Re$[\theta\f]$
signal ($\approx 2-3\times10^{-3}$ Rad, respectively), 
their contribution to Im$[\theta\f]$
was negligible.  The signals are linear in $B$, as expected.

Figure~\ref{fig;hallang} shows the temperature dependence of $\theta\h$
(a) and $\theta\h^{-1}$ (b) at 1079\cm\ and 8~T for a Au
sample.  The solid (empty)
circles represent the real (imaginary) part of $\theta\h$ and
$\theta\h^{-1}$. Note that Im[$\theta\h$] is greater than Re[$\theta\h$]
by approximately a factor of four, which suggests that the measurement is
approaching the high frequency regime where
Im[$\theta\h$]$>>$Re[$\theta\h$]. Though both Re[$\theta\h$] and
Im[$\theta\h$] show weak temperature dependence in
Fig.~\ref{fig;hallang}(a), only the Re[$\theta\h^{-1}$] shows temperature
dependence in Fig.~\ref{fig;hallang}(b).  This is consistent with a
temperature dependent $\gamma\h$ and temperature independent $\omega\h$
(see Eqs.~\ref{eq;hallang1} and \ref{eq;thetah2}).  The solid triangles in
Fig.~\ref{fig;hallang}(b) show the DC $\theta\h^{-1}$ at low and high 
temperatures.   The DC $R\h$ measurements 
were made using the Van der Pauw
geometry on the same
thin film samples that were examined in the MIR.
The DC $\theta\h^{-1}$ is seen to agree well with the MIR Re[$\theta\h^{-1}$].

Figure~\ref{fig;hallang2} shows the frequency dependence of
$\theta\h^{-1}$ at 290~K and 8~T. Im[$\theta\h^{-1}$] is represented by
empty circles which show a linear increase with frequency, suggesting a
frequency independent $\omega\h$ at these frequencies (see
Eq.~\ref{eq;thetah2}).  The Im[$\theta\h^{-1}$] data are fitted with a line
intersecting the origin, whose slope agrees with band predictions to
within 2~\%.  The Re[$\theta\h^{-1}$] is represented by solid circles and
shows no frequency dependence.  The DC $\theta\h^{-1}$ is also seen to
agree well with the MIR values.

Figure~\ref{fig;hallscat} shows the temperature dependence of the Hall
frequency $\omega\h$ and Hall scattering rate $\gamma\h$ at 
8~T. $\omega\h$ shows no temperature dependence in
Fig.~\ref{fig;hallscat}(a) and agrees well with the value obtained using
band theory (solid line).\cite{schulz,papa} Figure~\ref{fig;hallscat}(b)
shows the MIR $\gamma\h$ at 8~T (solid circles), the MIR $\gamma\xx$ (empty
squares), and the DC $\gamma\xx$ (empty triangles).  The MIR $\gamma\xx$ is
obtained from zero-field transmittance spectra (1000-7000\cm).  These
spectra are measured using a Fourier transform spectrometer and
are fitted with simple Drude theory to obtain the MIR $\sigma\xx$ and
$\gamma\xx$, which is used to transform $\theta\f$ into $\theta\h$.
 The DC $\gamma\xx$ are measured using conventional four probe electrical
techniques. Note that $\gamma\h$ is consistently smaller than the MIR
$\gamma\xx$ for the Au and Cu samples.  The MIR values for $\gamma\xx$ are
consistently higher than those obtained using DC measurements, especially
at lower temperatures.  Though exhibiting a similar temperature
dependence, the $\gamma\h$ are signifantly lower than both the MIR and DC
$\gamma\xx$.  Note that the scattering rates only decrease by 15-30~\% at
low temperature, which suggests that the scattering rate is dominated by
the interfaces of the films (as is discussed in Section~\ref{sec;aniso}).

Table~\ref{table1} shows the Hall scattering rates $\gamma\h$ and
$\gamma\xy$ along with the zero field scattering rate $\gamma\xx$.
Au films show a greater difference between $\gamma\xy$ and $\gamma\xx$
than the Cu film, with $\gamma\xy/\gamma\xx\approx 0.77$ for the
former and 0.86 for the latter.

Table~\ref{table2} shows the measured and predicted values for the Hall
frequency. $\omega\h$ agrees well with predictions from band
calculations.\cite{schulz,papa} For Cu, the measured $\omega\h$ is higher
than, but within 10~\% of the predicted\cite{schulz,papa} Hall frequency
$\omega\h^{\mathrm{band}}$.  For Au the agreement is even better, with
a variation of less than 5~\%.  The linear fit in
Fig.~\ref{fig;hallang2}(a) produces a measured $\omega\h$ that is within
2~\% of the band calculated value for Au.

Table~\ref{table3} shows the Hall 
coefficient $R\h=\sigma\xy/\sigma\xx^2$, from this and other
experiments.    $R\h^{\mathrm{band}}$
is derived from band calculations\cite{schulz} while $R\h^{\mathrm{free}}$ 
assumes a spherical FS and reduces to the well-known formula for
free electrons $R\h^{\mathrm{free}}=-1/nec$, where $n$ is the electron
density and $c$ is the speed of light.  $R\h^{\mathrm{bulk}}$ are 
from room temperature DC
measurements on bulk samples.\cite{hurd} Note that the MIR values for
Re[$R\h$] agree well with $R\h^{\mathrm{band}}$ for both Au and Cu.  
The MIR Im[$R\h$] are approximately a factor of five to eight times smaller
than Re[$R\h$] (see Section~\ref{sec;irdc}).  The DC $R\h$ for 
Au agrees well with $R\h^{\mathrm{free}}$ while the DC $R\h$  for 
Cu agrees better with the MIR measured value than
$R\h^{\mathrm{free}}$.  The MIR and DC measurements in both samples show little
temperature dependence.

\section{Discussion}
\label{sec;discuss}

\subsection{Anisotropic Fermi Surface Model}
\label{sec;aniso}

The difference between $\gamma\h$ and $\gamma\xx$ is a consequence of
anisotropic scattering on the FS.  This is seen more clearly from
Eqs.~\ref{eq;gammaxx} and \ref{eq;gammaxy} for $\gamma\xx$ and $\gamma\xy$.  
These quantities correspond to
two different averages of the scattering over the FS.  
From Eq.~\ref{eq;gammaxy} it is seen that $\gamma\xy$ weights more heavily
the regions of the FS with strong curvature.

A simple argument based on the FS anisotropy of Au and Cu can account for
the difference between $\gamma\xy$ and $\gamma\xx$. The
anisotropy of the FS in Au and Cu is well-characterized.  The high
curvature regions near the $L$ point of the FS are referred to
as necks while the low curvature regions everywhere else are referred to
as bellies. High curvature necks have a larger Hall conductivity
$\sigma\xy$ (see Eq.~\ref{eq;fsb}) while the lower curvature bellies, 
which make up a greater
fraction of the FS, tend to dominate the longitudinal conductivity
$\sigma\xx$.\cite{zhel,ioffe} Furthermore, the scattering rate of
carriers in the necks can be different from those in the bellies, with 
$\gamma\xy$ ($\gamma\xx$) tending to represent the 
characteristic scattering rate for carriers in the neck (belly).  In bulk
materials, the anisotropy in the scattering rate is due to the anisotropy
of the electron-phonon interaction.\cite{vfau1,vfau2}  However, in thin 
films the scattering
is dominated by the film interfaces and different mechanisms are responsible
for the anisotropy.  We explore one such mechanism, the anisotropy of the
Fermi velocity $v\f$, in the rest of this Section.

Since the bulk scattering length $l_0$ of the Au and Cu films 
at room temperature is roughly a
factor of four greater than the thickness of these films, one expects the
scattering length $l_{\mathrm{f}}$ in the films to be dominated by interface scattering.  
The ratio of the film conductivity $\sigma_{\mathrm{f}}$ and bulk conductivity
$\sigma_0$ is given by:\cite{ziman}

\begin{equation}
{\sigma_{\mathrm{f}}\over{\sigma_0}}\approx {3\over 4} {d\over {l_0}}
\ln\biggl({l_0\over d}\biggr)
\label{eq;ziman}
\end{equation}

\noindent where $d$ is the film thickness.  The conductivity ratio that
is measured in the Au (Cu) film is within 10~\% (30~\%) of the value
predicted by Eq.~\ref{eq;ziman}, and suggests that the scattering length
is dominated by the film thickness rather than impurities or phonons.  
Since the film consists of randomly oriented grains, and since $l_{\mathrm{f}}$
is independent of $k$, one expects the
average scattering length to be isotropic along the FS and related only to
the separation $d$ between film interfaces as implied by
Eq.~\ref{eq;ziman}.  For simplicity, we take the scattering time $\tau$ to
be related to the scattering length $l_{\mathrm{f}}$ and Fermi velocity $v\f$ as
follows:

\begin{equation}
l=v\f \tau\approx d
\label{eq;lk}
\end{equation}

\noindent
The scattering rate $\gamma$ can then be expressed as:

\begin{equation}
\gamma(k)=\frac{1}{\tau(k)}=\frac{v\f(k)}{l_{\mathrm{f}}}
\label{eq;gamma}
\end{equation}

\noindent where $\gamma(k)$ and $\tau(k)$ are the scattering rate and
scattering time of the carriers on the FS.  From this simple argument, one
can estimate that $\gamma(\hbox{neck})$ is related to
$\gamma(\hbox{belly})$ as follows:

\begin{equation}
\frac{\gamma\xy}{\gamma\xx}\rightarrow
\frac{\gamma(\hbox{neck})}{\gamma(\hbox{belly})}
=\frac{v\f(\hbox{neck})}{v\f(\hbox{belly})}\approx 0.56\ \hbox{to}\ 0.65
\label{eq;gammaratio}
\end{equation}

\noindent where the values for $v\f(\hbox{neck})$ and $v\f(\hbox{belly})$
represent the typical extrema of $v\f$ at the necks and bellies reported
in Refs.~\onlinecite{vfau1} and \onlinecite{vfau2}. Since the Hall 
measurements involve contributions 
from both neck and belly regions, the anisotropy 
observed in the experiment should be
smaller than that predicted by Eq.~\ref{eq;gammaratio},
which only involves extremal values of $v\f$.  
This can be seen in Table~\ref{table1}.
In fact, the results in Eq.~\ref{eq;gammaratio}
represent an upper limit of the anisotropy.  For a more
quantitative theoretical comparison, the FS integrals in Eqs.~\ref{eq;gammaxx} and 
\ref{eq;gammaxy} need to be calculated.

\subsection{Comparison of MIR and DC results}
\label{sec;irdc}

The frequency dependence of the values measured in this experiment provides 
important information about anisotropy and inelastic scattering.  In this Section,
we will discuss the frequency dependence of $R\h$, $\theta\h^{-1}$ and 
$\gamma\xx$. 
The Hall coefficient $R\h$ is shown in Table~\ref{table3}.  For Au, the MIR
value of Re[$R\h$] is independent of frequency and temperature and 20~\%
lower than the DC value.  Similar results are found for Cu, but with the 
low temperature MIR $R\h$  only 10~\% smaller than the DC value.  
This difference between the DC and MIR values for $R\h$ implies anisotropic
scattering on the FS (see Appendix~\ref{app}).  The MIR values of $R\h$ for Au and Cu
are in good agreement with
band calculations.\cite{schulz} On the other hand the DC value for Au
for these films (in contrast to the DC $R\h^{\mathrm{bulk}}$) gives the
correct carrier density, presumably fortuitously.  The MIR Im[$R\h$] is non-zero
for both Au and Cu, with a magnitude of 20~\% and 12~\% of the $R\h$, respectively. 
Non-zero Im[$R\h$] occurs only for a frequency dependent Re[$R\h$], and 
the magnitude of Im[$R\h$] is related to the difference between the DC and MIR
Re[$R\h$].
Therefore, the larger
Im[$R\h$] in Au is consistent with the stronger 
frequency dependence of Re[$R\h$] in Au when compared to Cu. 
The sign of Im[$R\h$] is
consistent with the relative magnitudes of $\gamma\xy$ and $\gamma\xx$ 
(see Eqs.~\ref{eq;rh2} and \ref{eq;gr}), as found from a comparison of
$\sigma\xx$ with $\theta\h$.  Indeed, $\gamma\xx\ne\gamma\xy$ is due
to anisotropic scattering and gives the most revealing information about
the anistropy of the scattering.  

In the case of the inverse Hall angle the DC value $\theta\h^{-1}$ is in good
agreement with Re[$\theta\h^{-1}$] in the MIR.  However, this is also
fortuitous since the asymptotic expression Eq.~\ref{eq;thetah2} is not
valid at low frequency because of the anisotropy of the scattering on the
FS.  Moreover, even for isotropic scattering, the AC value of
Re[$\theta\h^{-1}$] should be larger because of phonon scattering at high
frequencies.  Therefore it appears that the anisotropy effects and the
inelastic scattering effects nearly cancel in this case.

From $\sigma\xx$ we deduce that the DC $\gamma\xx$ is smaller than the
IR $\gamma\xx$. This is expected both from anisotropy of the scattering
and its frequency dependence. The effect of anisotropy follows from the general result
that $\langle1/\tau\rangle\langle\tau\rangle\ge 1$ for 
averages over the FS, since $\rho_{\mathrm{DC}}\sim\langle\tau\rangle^{-1}$ 
(Eq~\ref{eq;fsa}) and $\gamma\xx\sim\langle\tau^{-1}\rangle$ (Eq~\ref{eq;gammaxx}). 
The effect of anisotropic scattering should be comparable to that observed in the Hall
coefficient or the difference between $\gamma\xx$ and $\gamma\xy$.  
Enhanced electron-phonon scattering in the MIR is also expected due to phonon
emission at low temperatures which cannot occur for DC excitation.  The phonon 
emission component to the MIR scattering rate is estimated to be 
approximately 5~\%.\cite{phonon}
The frequency dependence due to anisotropy and electron-phonon 
interaction is quantitatively
consistent with the difference between  the MIR and DC $\gamma\xx$.

\section{Conclusion}

\label{sec;conclusion}

We have demonstrated a sensitive MIR photo-elastic polarization modulation technique
that can be used to obtain the complex MIR Hall conductivity in thin-film metals.
The Hall frequencies obtained from these measurements are consistent with
band theory, while the Hall scattering rates are consistently lower than
those predicted from Drude analysis of zero magnetic field transmission
measurements. This difference can be explained qualitatively in terms of
the difference in Fermi velocities, and hence the difference in scattering
rates, for carriers in the neck and belly regions of the FS. The neck and
belly regions contribute differently to $\sigma\xx$ and $\sigma\xy$,
which can account for the difference in $\gamma\h$ and $\gamma\xx$.
We hope that further band structure calculations will be made to better
quantify these arguments.  MIR Hall angle measurements have provided a
sensitive probe of the FS anisotropy in thin-film metals, and since a
number of theories\cite{zhel,ioffe} predict an anisotropic FS in high
temperature superconductors, this technique may be useful for studying
these less conventional materials.


\acknowledgments

This work was supported in part by NSF grant DMR-9705129 and by funding from the NSA.

\appendix
\section{THEORETICAL BACKGROUND FOR THE AC HALL EFFECT}
\label{app}

In this Appendix, we discuss the magneto-conductivity tensor within
Fermi Liquid theory\cite{landau} and the reelaxation time approximation. These 
results provide a useful basis for analyzing the experimental data. 
Applying the small angle approximation $\tan\theta\h\approx\theta\h$, 
the sum rule on $\theta\h$ is\cite{drew}

\begin{equation}
\int^\infty_0 \hbox{Re[}\theta\h\hbox{]}d\omega =
\frac{\pi}{2}\omega\h. 
\label{eq;sum}
\end{equation}

Since $\omega$ is greater than the
carrier relaxation rates in the experiment discussed in this paper, 
it is useful to consider the asymptotic
forms of the magneto-optical response functions for large $\omega\tau$.
For large $\omega\tau$

\begin{equation}
\tilde\tau(k)=\frac{1}{\frac{1}{\tau(k)}-i\omega}
=\frac{i}{\omega}+\frac{\gamma(k)}{\omega^2}+\cdots,
\end{equation}

\noindent where $\gamma(k)=1/\tau(k)$ is the $k$-dependent scattering rate on
the FS.  The limiting high frequency behavior of the conductivity tensor in
Eqs.~\ref{eq;fsa} and \ref{eq;fsb} becomes

\begin{mathletters}
\begin{eqnarray}
\sigma\xx=&&\sigma\xx^{\infty}\biggl(1-i\frac{\gamma\xx}{\omega}+\cdots\biggr)
\label{eq;sigmaxx}\\   
\sigma\xy=&&\sigma\xy^{\infty}\biggl(1-2i\frac{\gamma\xy}{\omega}+\cdots\biggr).
\label{eq;sigmaxy}   
\end{eqnarray}
\end{mathletters}

\noindent $\gamma\xx$ and $\gamma\xy$ are different averages of the
scattering over the FS given by:

\begin{mathletters}
\begin{eqnarray}
\gamma\xx=&& \frac{\oint_{\hbox{\tiny FS}}dS\
|{\bf v}(k)|\gamma(k)}{\oint_{\hbox{\tiny FS}}dS\ |{\bf v}(k)|}
\label{eq;gammaxx} \\
\gamma\xy=&& \frac{\oint_{\hbox{\tiny FS}}dS\
\gamma(k){\bf e_z}\cdot\biggl[{\bf v}(k)\times 
{ {d\ {\bf v}(k)}\over{dk}}\biggr]}{\oint_{\hbox{\tiny FS}}dS\
{\bf e_z}\cdot\biggl[{\bf v}(k)\times 
{ {d\ {\bf v}(k)}\over{dk}}\biggr]}.
\label{eq;gammaxy}
\end{eqnarray}
\end{mathletters}

\noindent $\sigma\xx^{\infty}$ and $\sigma\xy^{\infty}$ are given by:

\begin{mathletters}
\begin{eqnarray}
\sigma\xx^{\infty}=&& \frac{i\omega_p^2}{4\pi\omega}
\label{eq;limitxx}\\   
\sigma\xy^{\infty}=&& -\frac{\omega_p^2\omega\h}{4\pi\omega^2},
\label{eq;limitxy}
\end{eqnarray}
\end{mathletters}

\noindent where $\omega_p$ and $\omega\h$ are the plasma frequency and the
Hall frequency, respectively, given by
integrals over the FS defined below where we have assumed cubic symmetry and 
used the weak field approximation.

\begin{eqnarray}
\omega_p^2=&&  \frac{e^2}{2\pi^2\hbar}
\oint_{\hbox{\tiny FS}} dS\ |{\bf v}|\\
\label{eq;wp}
\omega\h=&& \frac {eH}{\hbar c}
\frac{\oint_{\hbox{\tiny FS}} dS\ {\bf e_z}
\cdot\biggl[{\bf v}\times\frac{d{\bf v}}{dk}\biggr]}
{\oint_{\hbox{\tiny FS}} dS\ |{\bf v}|}
\label{eq;wh}
\end{eqnarray}

\noindent Consequently, the asymptotic forms of $\theta\h$ and the Hall coefficient
$R\h$ are

\begin{eqnarray}
\theta\h^{\infty}=&& \frac{i\omega\h}{\omega}\\
\label{eq;thetah}   
R\h^{\infty}=&& \frac{4\pi\omega\h}{\omega_p^2}
\label{eq;rh}
\end{eqnarray}

\noindent Keeping the first order terms in $\tilde\tau$ in 
Eqs.~\ref{eq;sigmaxx} and \ref{eq;sigmaxy}, one obtains the
asymptotic expansions for the inverse Hall angle $\theta\h^{-1}$ and Hall
coefficient $R\h$

\begin{eqnarray}
\theta\h^{-1}=&& \frac{-i\omega}{\omega\h}+\frac{\gamma\h}{\omega\h}+
O\biggl(\frac{i}{\omega}\biggr)
+O\biggl(\frac{1}{\omega^2}\biggr)+\cdots
\label{eq;thetah2} \\ 
R\h=&& \frac{\sigma\xy}{\sigma\xx^2}
=\frac{4\pi\omega\h}{\omega_p^2 B}\biggl[1+ \frac{i}{\omega}\gamma_R+
O\biggl(\frac{1}{\omega^2}\biggr)
+O\biggl(\frac{i}{\omega^3}\biggr)+\cdots\biggr]
\label{eq;rh2}
\end{eqnarray}

\noindent where $B$ is the magnetic field and

\begin{mathletters}
\begin{eqnarray}
\gamma\h\equiv && 2\gamma\xy-\gamma\xx \label{eq;gh} \\
\gamma_R\equiv && 2(\gamma\xx-\gamma\xy)\label{eq;gr}.
\end{eqnarray}
\end{mathletters}

Equation~\ref{eq;thetah2} for $\theta\h^{-1}$ is particularly useful since
the scattering effects and $\omega\h$ can be readily separated.  Furthermore, 
the high frequency asymptotic form is seen to reduce to the simple Drude form of
Eq.~\ref{eq;hallang1}.  Also, it is seen that $R\h$ becomes frequency
dependent only when the scattering rate is not constant on the FS.
This follows from either Eq.~\ref{eq;rh2} where the leading frequency
dependent term is proportional to $\gamma_R=2(\gamma\xx-\gamma\xy)$ (which is zero
for isotropic scattering), or in 
Eqs.~\ref{eq;fsa} and \ref{eq;fsb} where a $k$-independent $\gamma$ 
can be taken out of the integrals and exactly cancels in the ratio 
$\sigma\xy/\sigma\xx^2$.

These expressions can be extended to include frequency-dependent, 
inelastic scattering in the
memory function formalism.\cite{allen}  In this 
case $\gamma(k)-i\omega$ is replaced
with $\Gamma(k,\omega)-i\Sigma(k,\omega)-i\omega=
\Gamma(k,\omega)-i\omega(1+\lambda(k,\omega))$,
where $\Gamma$ and $\Sigma$ are the real and imaginary parts of the memory
function, respectively.  In this case simple expressions can be obtained
only for $k$-independent scattering.  It is seen that the Hall coefficient
is unaffected since the scattering effects cancel. Therefore, $R\h$
remains frequency independent.  The result for $\theta\h^{-1}$ is

\begin{equation}
\theta\h^{-1}=\frac{\Gamma(\omega)-i\omega(1+\lambda(\omega))}{\omega\h} 
\label{eq;mem}
\end{equation}

Therefore, similar to the behavior of $\sigma\xx$, the Hall angle has the same form
as in the elastic scattering case but with renormalized parameters.  The
real part of $\theta\h^{-1}$ gives the scattering function while the
imaginary part gives the renormalized Hall frequency
$\omega\h^*=\omega\h/(1+\lambda(\omega))$.  The Hall angle sum rule is
satisfied because $\lambda(\omega)\rightarrow 0$ as
$\omega\rightarrow\infty$.

\begin{figure}[tbp]
\caption{A schematic of the experimental setup.}
\label{fig;setup}
\end{figure}

\begin{figure}[tbp]
\caption{The complex Faraday angle $\theta\f$ for a Cu film
as a function of magnetic field at 949\cm\  and 290~K.}
\label{fig;signal}
\end{figure}

\begin{figure}[tbp]
\caption{The complex Hall angle $\theta\h$ (a) and inverse Hall angle
$\theta\h^{-1}$ (b) for Au as a function of temperature at 1079\cm\ and
8~T. The solid (empty) circles correspond to the real (imaginary) part of $\theta\h$ 
and $\theta\h^{-1}$ in the MIR.  The solid triangles
in (b) show the DC values for $\theta\h^{-1}$.}
\label{fig;hallang}
\end{figure}

\begin{figure}[tbp]
\caption{The complex inverse Hall angle $\theta\h^{-1}$ for Au is shown as a
function of frequency at 290~K and 8~T.  The solid (empty) circles represent the 
real (imaginary) part
of $\theta\h^{-1}$.  Re[$\theta\h^{-1}$] shows no frequency dependence, while  
the empty circles representing
Im[$\theta\h^{-1}$] shows a linear temperature dependence as expected in
Eq.~\protect\ref{eq;thetah2}.} 
\label{fig;hallang2}
\end{figure}

\begin{figure}[tbp]  
\caption{The Hall frequency $\omega\h$ (a) and Hall scattering rate 
$\gamma\h $ (b) for Au as a function of temperature at 1079\cm\ and 8~T.
$\omega\h$ is independent of temperature and agrees well with the 
prediction from band calculations, which is shown
by the solid line in (a). MIR $\gamma\h$ at 8~T (solid circles), the  MIR $\gamma\xx$ 
(empty squares), and the DC $\gamma\xx$ (empty triangles) are shown in (b). 
Despite large differences in magnitude, these
scattering rates show similar temperature dependence.}
\label{fig;hallscat}
\end{figure}

\begin{table}[tbp]
\caption{Comparison of the Hall scattering rates $\gamma\h$ and $\gamma\xy$ 
(see Eq.~\protect\ref{eq;gh}) obtained from
magneto-optic measurements with the longitudinal scattering
rate $\gamma\xx$
obtained from a Drude fit to zero magnetic field transmission measurements.
Note the strong anisotropy between the $\gamma\h$ and $\gamma\xx$.
All measurements were performed at 290~K.}
\label{table1}
\begin{tabular}{llllll}
Sample & Frequency (\cm) & $\gamma\h$ (\cm) & $\gamma\xy$ (\cm) 
& $\gamma\xx$ (\cm) & $\gamma\xy$/$\gamma\xx$ \\ 
\tableline 
Cu & 949 & 525\ $\pm$\ 55 & 605\ $\pm$\ 60 & 685\ $\pm$\ 70 & 0.88 \\ 
Au & 1079 & 449$\ \pm$\ 45 & 585\ $\pm$\ 60 & 720\ $\pm$\ 76 & 0.81 \\
\end{tabular}
\end{table}

\begin{table}[tbp]
\caption{Comparison of the Hall frequency $\omega\h$ with the 
$\omega\h^{\mathrm{band}}$ predicted from band
calculations.\protect\cite{schulz,papa}  The agreement is 
within the experimental error.
All measurements were performed at 290~K.}
\label{table2}
\begin{tabular}{lllll}
Sample & Frequency (\cm) & $\omega\h$ (\cm) & $\omega\h^{\mathrm{band}}$ (\cm)
& 
$\omega\h $/$\omega\h^{\mathrm{band}}$ \\ 
\tableline 
Cu & 949 & -4.1\ $\pm$\ 0.2 & -3.8 & 1.08 \\ 
Au & 923 & -5.6\ $\pm$\ 0.3 & -5.5 & 1.02\\
Au & 1079 & -5.2\ $\pm$\ 0.3 & -5.5 & 0.95\\
\end{tabular}
\end{table}

\begin{table}[tbp]
\caption{Comparison of the MIR and DC Hall coefficients $R\h$. The
units for $R\h$ are $10^{-11}$~m$^3$C$^{-1}$. The Cu and Au MIR measurements
were performed at 949\cm\ and 1079\cm, respectively.}
\label{table3}
\begin{tabular}{lllllll}
Sample & T(K) 
& MIR $R\h$ & DC $R\h$ 
& $R\h^{\mathrm{band}}$ Ref.~\onlinecite{schulz}
& $R\h^{\mathrm{free}}$ Ref.~\onlinecite{schulz} 
& DC $R\h^{\mathrm{bulk}}$ Ref.~\onlinecite{hurd}\\
\tableline
Cu & 290  & -4.42\ -0.465i & -4.28  & -5.2   &-7.3 & -5.17    \\
Cu & 40   & -4.28\ -0.534i  & -4.75  &        &     &        \\   
Au & 290  & -8.04\ -1.49i  & -9.9   &  -8.1 &-10.5 & -7.16  \\
Au & 23   & -7.96\ -1.65i  & -10.4  &        &     &        \\    
\end{tabular}
\end{table}


\begin{references}
\bibitem{general}The generalization to lower symmetry is given in 
Refs.~\protect\onlinecite{ong} and \protect\onlinecite{zhel}.
\end{references}
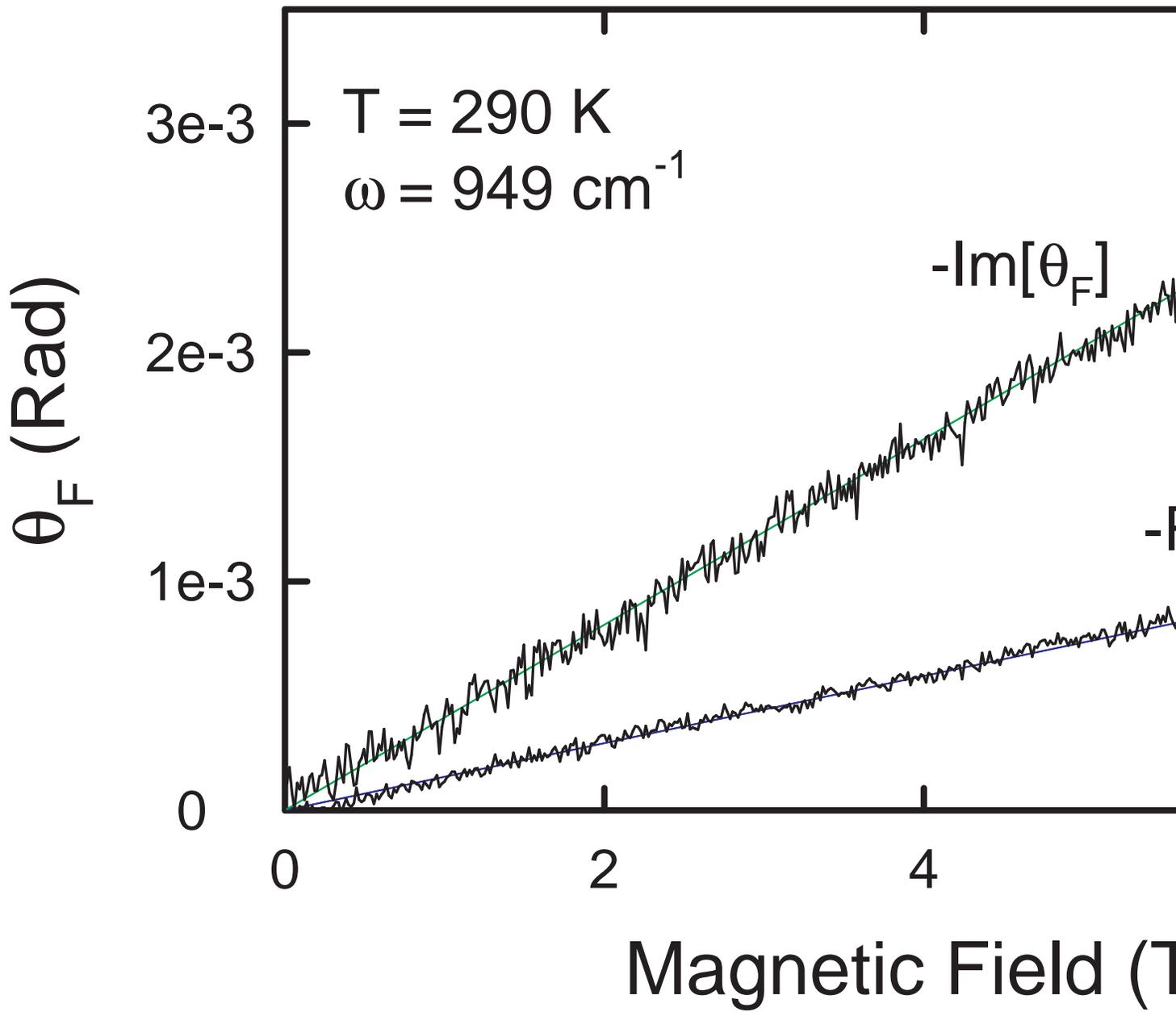
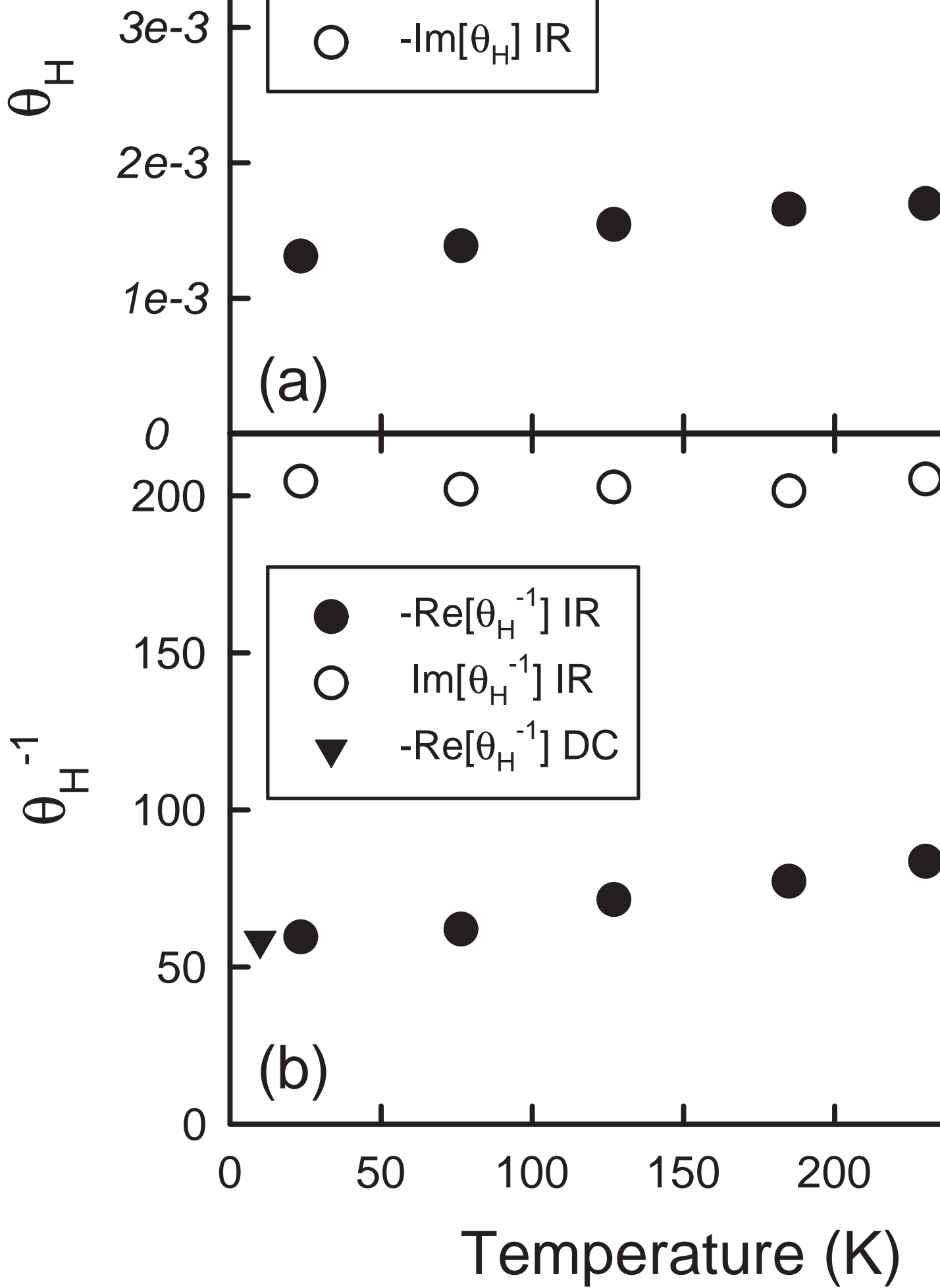
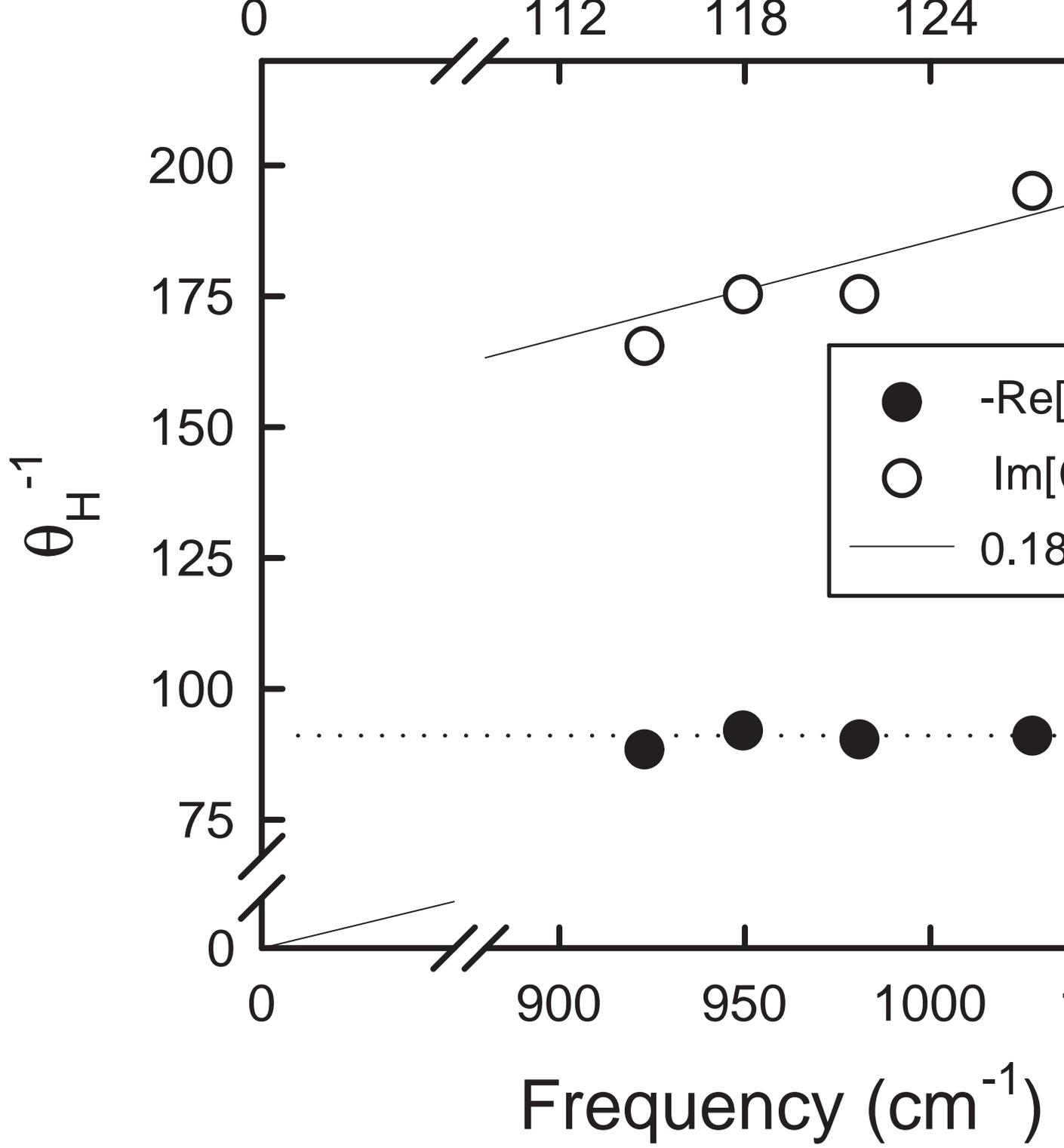
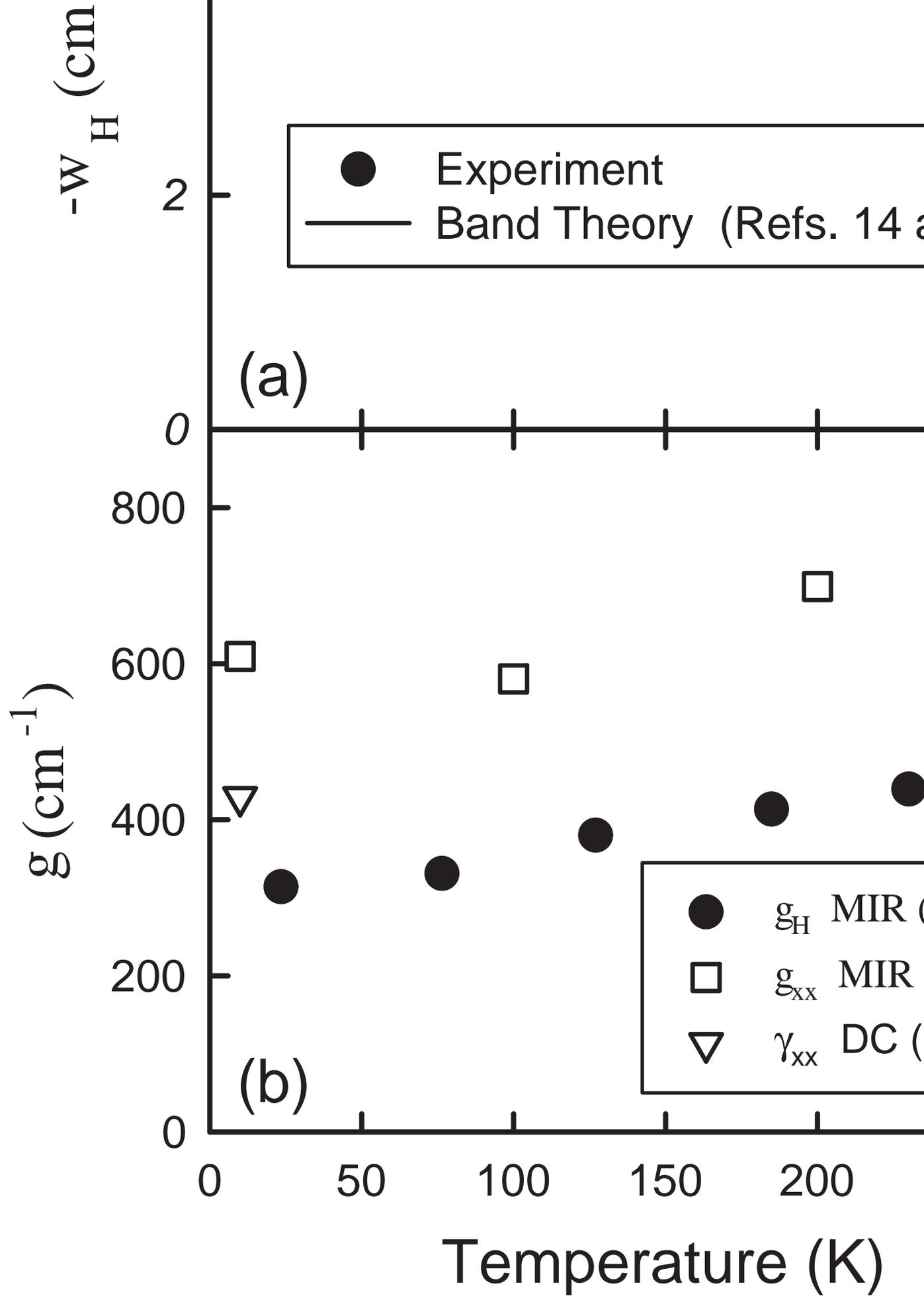
\end{document}